# Narcotweets: Social Media in Wartime

**Andrés Monroy-Hernández, Emre Kiciman, danah boyd, Scott Counts**
Microsoft Research
{andresmh, emrek, dmb, counts}@microsoft.com

**Abstract**

This paper describes how people living in armed conflict environments use social media as a participatory news platform, in lieu of damaged state and media apparatuses. We investigate this by analyzing the microblogging practices of Mexican citizens whose everyday life is affected by the Drug War. We provide a descriptive analysis of the phenomenon, combining content and quantitative Twitter data analyses. We focus on three interrelated phenomena: general participation patterns of ordinary citizens, the emergence and role of information curators, and the tension between governmental regulation and drug cartel intimidation. This study reveals the complex tensions among citizens, media actors, and the government in light of large scale organized crime.

## Introduction

Social media have emerged as a communication channel people use to connect with others and get information in extraordinary moments of crisis: from floods (Bruns, Burgess, Crawford, & Shaw, 2012; Starbird, Palen, Hughes, & Vieweg, 2010) to earthquakes (Starbird & Palen, 2011) to terrorist attacks (Cheong & Lee, 2010) to school shootings (Palen & Vieweg, 2008) to revolutions (Al-Ani, Mark, Chung, & Jones, 2012; Lotan et al., 2011). However, *in localities afflicted with armed conflicts, crises are part of everyday life*, turning otherwise extraordinary events into ordinary ones. Additionally, risk has a more important role. For example, what otherwise would be controversial speech, during times of war it can easily put people's lives at risk. As we will discuss, people living in these circumstances can turn to social media to meet their information needs, especially when the institutions traditionally charged with providing information are undermined. Previous work showed that social media and other forms of communication technology in wartime helps citizens maintain routine, enable discussion of topics that were socially unacceptable to discuss face to face, and even change social practices such as finding marriage partners (Mark & Semaan, 2008).

In this paper, we examine Twitter use in a different war: the Mexican Drug War, an ongoing armed conflict in a country with more than thirty million Internet users and a highly visible presence of social media in public life, which, for many, has turned into a fluid and participatory information platform. Our goal is to document how citizens are using social media as a resource for both practical purposes and social support.

We focus on three interrelated phenomena: general participation patterns of ordinary citizens, the emergence and role of information curators, and the tension between governmental regulation and drug cartel intimidation.

## Background and Related Work

The Mexican Drug War is an ongoing conflict of such a large scale that, from 2006 to 2011, it took the lives of more than sixty thousand people (10% of them civilians), and has displaced between 230,000 and 1.6 million people (Internal Displacement Monitoring Centre, 2011, "Quinto año de gobierno: 60 mil 420 ejecuciones," 2011). The frequent clashes among drug cartels, the military, and police forces often take place in urban areas, inducing panic and cases of post-traumatic stress disorder (Menchaca, 2011).

Like other armed conflicts, the Mexican Drug War is also a conflict over the control of information flow. To date, the drug cartels seem to be winning control of mainstream news media through murder and intimidation. Mexico is now ranked as the third most dangerous country for journalists (Committee to Protect Journalists, 2011). The result is a "near-complete news blackout" imposed by criminal organizations who "via daily telephone calls, e-mails and news releases" decide "what can and cannot be printed or aired" (Booth, 2010). Local governments, likewise, often fail in public communication for a variety of reasons: fear of reprisals, lack of knowledge, or an attempt to maintain an image of having everything under control. For example, the foreign press reported on how a mayor of a northern city "mysteriously disappears for days and refuses to discuss drug violence" (Booth, 2010).

Some newspapers have publicly vowed not to report on the violence, for example, the leading newspaper in Saltillo -one of the cities we analyze in this paper- published an editorial announcing that because of the "threats on their editorial staff"[1] they were "obligated, on occasion, to leave out information" ("Entre sombras," 2010).

---
[1] This and other quotes are translated from Spanish.

This "information vacuum" is filled by social media. As the drug cartels succeed in intimidating traditional news media, the battle for the control of information is shifting to social media. This shift decentralizes the flow of information, leading to issues with anonymity and trust in information sources, and as we will show, it creates a role for what Castells (2009, pp. 362–363) calls "insurgent communities" that "emerge from network individuals reacting to perceived oppression, then transforming their shared protest into a community of practice", which in Mexico also leads to a new information hub in the social media ecosystem: the curator.

### Social Media in Mexico

Parallel to the increase in violence, Mexico has also seen an increase in the adoption and visibility of the Internet and social media in public life. From 2000 to 2010 the number of Mexicans with access to the Internet increased from 17.2% to 34.9% (Asociación Mexicana de Internet, 2011). In 2010, there were more than 34 million Mexicans with access to the Internet. Partly because of the demographics of the country, most Mexicans online are young: 37% of people online are 17 or younger, and 40% are between 18 and 34. Additionally, the majority of people with access to the Internet (61%) participate on social media sites, primarily Facebook (39%), YouTube (28%), and Twitter (20%). The majority of users (53%) access Twitter at least once a day, from home (39%), work (16%), or mobile phone (18%). Thus, besides filling the void of traditional news sources, these numbers underscore the practicality and effectiveness of social media as an information channel that can reach a significant percentage of the population in a timely manner.

### Social Media as Civic Media

Frustrated by the lack of news, citizens have turned to social media. Twitter in particular has come to be one of the principal sources of citizen-driven news in Mexico. People often report, confirm, spread, and comment on information about violent events using keywords known as hashtags that have emerged as shared news resources. Many Mexicans see social media as a source of "information and survival" (Cave, 2011). People also use social media to collectively grieve, and express frustration toward the government, the criminal organizations, and themselves for the circumstances in which they live.

Ethnographic research on the Gulf War showed that Iraqis have done the same with blogging: "[t]he majority of bloggers who lived in Iraq reported events as they unfolded, as eyewitnesses" (Al-Ani, Mark, & Semaan, 2010). Blogs also provided a counter narrative to government-controlled media of events in the recent Egyptian revolution (Al-Ani et al., 2012). Use of social media in the Mexican Drug War incorporates aspects of such counter-narrative while also highlighting the practical uses borne out of the necessity of crisis scenarios. Finally, blogs have been shown to reflect important events in continuing struggles (Mark et al., 2012)

### Participation Patterns

Much has been said on the media about Mexican citizens using Twitter to report violent events in their communities. However, little is known about how the volume of tweets and of users involved. Are there just a few people posting most of the content? What proportion of the content are retweets as opposed to *de novo* tweets? Through an examination of tweets on the drug war in Mexico, we show that people use Twitter to spread information and present an overview of Twitter participation patterns that highlight, among other things, how people collectively describe and document important events, and how subcommunities emerge through use of the hashtag convention.

We focused on four Mexican cities based on their prominence in the national and international news media, our familiarity with their social and cultural context, and their diversity in size and location. Leveraging our observations of the Mexican social and traditional media, we identified a set of hashtags commonly used on Twitter to tag events linked with the Mexican Drug War in those (see Table 1). We used the Twitter "firehose", available to use through our company's contract with Twitter, to collect the entirety of tweets containing those words posted in the 16-month period between August 2010 and November 2011. This resulted in a corpus of 596,591 tweets posted.

| City | Hashtag | Daily Tweets | |
|---|---|---|---|
| | | Mean | Max |
| Monterrey | #mtyfollow | 471 | 7,724 |
| Reynosa | #reynosafollow | 337 | 2,806 |
| Saltillo | #saltillo | 660 | 7,105 |
| Veracruz | #verfollow | 650 | 4,985 |

Table 1: Daily Tweets per City

### Results

**Content.** We found that half (49.5%) of all posts associated with the four cities, were contributing new content (*de novo* tweets), while a third (30.7%) were spreading existing messages (retweets), and one-fifth (19.9%) were mentions and replies[2] to others.

Additionally, as expected, a significant number of the tweets were reports of continuing violent events that included their whereabouts. For example, in the Monterrey data set, the words related to locations --such as the

---

[2] References can be either mentions (for example, "they say that @mary is not posting accurate reports #city") or replies (for example, "@mary: I saw a shooting by First Avenue #city")

Spanish words for "avenue" and "south" -- amounted to at least 1% of all words (see Table 2); which, although small, represent the highest proportion of all words that were not stop words (namely, common words such as prepositions and articles). Interestingly, the hashtag for a different city (Reynosa) is also one of the most common words in the Monterrey data, indicating a link between the two cities that would explain the similarities in Twitter practices and expected social connections.

| Meaning | Probability | Original Tokens |
| --- | --- | --- |
| Locations within the city | 1.0% | zona, san, sur, altura, garza, col., av. |
| Shootings, Blasts | 0.7% | #balacera, balacera, balazos, detonaciones |
| Report | 0.3% | reportan |
| References to people | 0.3% | gente, alguien |
| Hashtag for Reynosa | 0.2% | #reynosafollow |

**Table 2: Most common words in Monterrey's tweets**

**Individual contributions.** We found 56,414 different users across the four cities, who, on average, posted 9.4 tweets each. More specifically, those who posted new reports did so at a rate of 10.8 per person, those who spread tweets did so at a rate of 4.1 retweets per person, and those who referenced others did so at a rate of 5.5 per person. A few users contributed most of the content. For example, in the hashtag associated with Monterrey, half the people (52.8% or 14,898) tweeted only once, while a tiny fraction (0.03% or 9 people) tweeted more than a thousand times each. This uneven distribution highlights the importance of the *curators*, whose role we discuss the next section.

### Discussion

Putting this data in the demographic context of the cities we examined, it is important to note that, on average, 1.48% of people living in the cities analyzed posted something on Twitter about the drug war. Assuming these cities have similar Internet penetration as the rest of Mexico (34.9% of the population, as cited above), an average of 4.2% of the online population has posted something about the drug wars on Twitter. These estimates have many limitations because they assume: 1) people tweeting with those hashtags live in the city associated with the hashtag, 2) Internet penetration is homogeneous, and 3) people have only one account. Conversely, not every person uses the hashtags we analyzed. As for the patterns of participation, they show the importance not only of posting reports but also of spreading those messages. Furthermore, the chatter among people might indicate the creation of bonds among those participating in the hashtags. The community aspect is exemplified by one of the most common tweets on the Reynosa hashtag being the greeting "good morning" to the members of a hashtag. In addition, abuse is chastised. For example, a tweet on #mtyfollow warns, "Sirs, I have created #mtyfollou so you can write your f*** stupidities there and stop misusing #mtyfollow."

### The Emergence of Civic Media Curators

The results of our first analysis showed that a few users posted much of the content. In this study, we describe who these central actors are. Our findings suggest that some of them represent citizen efforts intended to address the need for curation and aggregation of the rush of real-time tweets of acute events.

For this analysis, we focused only on the city of Monterrey. First, we generated a list of the users who have contributed the most to the city's tweets by counting the number of times each user posted a message with the city's hashtag.

We put together a list of the most active users[3] by selecting those individually responsible for 1% or more of the Monterrey's tweets (with the hashtag #mtyfollow). This resulted in a list of ten users: one was responsible for posting 3% of all tweets, two were responsible for 2%, and the remaining seven were each responsible for 1% of the posts. Some of those who posted frequently had few followers, while others had many followers (median: 9,305). Likewise, some accounts were regularly mentioned by others while others received few mentions (median: 303). The four accounts with more followers and mentions than the median all represented a standard practice that can be understood as *information curation*: receiving, responding to, and retweeting dozens of tweets from other account. Examining the dynamics of these accounts sheds light on the kinds of practices that are emerging in this ecosystem.

### Regulation and Intimidation

Given governmental anxiety around citizen curators, we decided to examine what happens when information flow is regulated through government interventions or even intimidation from the drug cartels. We present the events of Veracruz as a case study to examine the government reaction. The case highlighted the challenges with trust and government regulation.

On August 25, 2011 at 11:56 AM, a twitter user reported that five kids had been kidnapped at a school: "#verfollow I confirm that in the school 'Jorge Arroyo' in the Carranza neighborhood 5 kids were kidnapped by an armed group, there's panic in the zone." The message was retweeted by twelve people, including one of the curators of #verfollow, an account with more than 5,000 followers. Immediately, the rumor started to spread like wildfire. By noon, only four minutes after the first tweet, the governor of the state of Veracruz tweeted from his personal account that the rumor was false. However, by then it was either too late, or the governor was not considered a reliable news source. Dozens of parents rushed to pick up their children from school, causing massive traffic, chaos, and panic across the city. Three hours later, the governor tweeted that his government would go after those who spread the rumor and later the state government website made a statement listing sixteen twitter accounts involved in the rumor and threatening to take legal action against them. The statement also mentioned the full name of the people associated with one of the accounts. By Saturday, two people had been were arrested on charges of terrorism.

---
[3] We avoid mention the usernames of the people studied. We make exceptions for highly public accounts.

Many *tuiteros* countrywide rallied in opposition to the arrests, jokingly calling themselves "Twitteroristas." After much pressure, the state government released the twitter users but enacted a new law that penalizes the use of social network sites for spreading misinformation that might cause panic. Some argued that this new law and the imprisonment of those two twitter users would have a "chilling effect" on expression.

There has also been much speculation that criminal organizations have infiltrated social media to spread panic in the population. For example, Okeowo (2010) reports on how "drug cartels apparently use Twitter and Facebook not only to communicate with one another, but also to spread fear through local communities."

## Conclusion

We have shown that social media creates an alternate "user-generated" channel of communication that can address weaknesses in information flow. However, this new channel comes with its own challenges such as issues of trust, reputation, and misinformation.

Various solutions have shown their own weaknesses. On the one hand, government regulation has the potential of cooling free speech: fear of punishment can undermine people's willingness to contribute to a common good.

Additionally, there is a common refrain among proponents of social media in Mexico that argue that the government needs to use social media if it wants more control of information flow. However, as the case presented here show, it is not only a matter of being present but about engaging in fruitful interactions overtime that can help the government, just as anyone else, to gain reputation and creditability among social media users.